\documentclass[12pt]{article}
\usepackage{graphics}                                                      
                                                       
\newlength{\dinwidth}                                                       
\newlength{\dinmargin}                                                       
\def\lapproxeq{\lower .7ex\hbox{$\;\stackrel{\textstyle                                                       
<}{\sim}\;$}}                                                       
\def\gapproxeq{\lower .7ex\hbox{$\;\stackrel{\textstyle                                                       
>}{\sim}\;$}}                                                       
\def\be{\begin{equation}}                                                       
\def\ee{\end{equation}}                                                       
\def\bea{\begin{eqnarray}}                                                       
\def\eea{\end{eqnarray}}

\begin{document}                                                       
\begin{flushright}                                                       
IPPP/01/01 \\ 
DCPT/01/02 \\                                                       
10 January 2001 \\     
\end{flushright}                                                       
                                                       
\vspace*{2cm}                                                       
                                                       
\begin{center}                                                       
{\Large \bf Gluon shadowing in the low \boldmath{$x$} region probed by the LHC} \\

\vspace*{1cm}                                                       
M.A. Kimber$^a$, J. Kwiecinski$^{a,b}$ and A.D. Martin$^a$                                                        
                                                      
\vspace*{0.5cm}                 
$^a$ Department of Physics, University of Durham, Durham, DH1 3LE \\                                                      
$^b$ H.~Niewodniczanski Institute of Nuclear Physics, ul.~Radzikowskiego 152,                                    
Krakow, Poland \\          
\end{center}                                                       
                                                       
\vspace*{1cm}                                                       
                                                       
\begin{abstract}                                                       
Starting from an unintegrated gluon distribution which satisfies a `unified' equation  
which embodies both BFKL and DGLAP behaviour, we compute the shadowing corrections  
to the integrated gluon in the small $x$ domain that will be accessible at the LHC.  The  
corrections are calculated via the Kovchegov equation, which incorporates the leading $\ln  
(1/x)$ triple-Pomeron vertex, and are approximately resummed using a simple Pad\'{e}  
technique.  We find that the shadowing corrections to $xg (x, Q^2)$ are rather
small in the HERA domain, but lead to a factor
of 2 suppression in the region $x \sim 10^{-6}$, $Q^2 \sim 4~{\rm GeV}^2$
accessible to experiments at the LHC. 
\end{abstract}                                             
              
\vspace*{1cm} 
Experiments at HERA and the Tevatron have confirmed the rapid increase of the gluon  
distribution as $x$ decreases, which is expected both in the pure DGLAP framework and in  
the BFKL-motivated approach.  It is anticipated, however, that at sufficiently small $x$, this  
increase will be tamed by shadowing corrections. 
 
The first quantitative studies of gluon shadowing were made by Gribov, Levin and Ryskin  
\cite{GLR} (GLR) and by Mueller and Qiu \cite{MQ}.  It was found that the shadowing  
contribution, $f_{\rm shad} (Y, k^2)$, to the gluon distribution $f (Y, k^2)$, unintegrated  
over the gluon transverse momentum $k$, is of the form 
\be 
\label{eq:a1} 
\frac{\partial f_{\rm shad} (Y, k^2)}{\partial Y} \; = \; - C \: \frac{\alpha_S^2}{R^2} \:  
\frac{1}{k^2} \: \left [ xg (x, k^2) \right ]^2, 
\ee 
where $Y \equiv \ln (1/x)$, $\pi R^2$ denotes the transverse area populated by the gluons and  
$g$ is the integrated gluon distribution.  The constant $C$ will be specified later.  When the  
shadowing term is combined with DGLAP evolution in the double leading $\log$  
approximation (DLLA) then we obtain the GLR equation for the integrated gluon 
at scale $k^2$
\be 
\label{eq:a2} 
\frac{\partial g (x, k^2)}{\partial Y \: \partial \ln (k^2/\Lambda^2)} \; = \; \frac{N_C  
\alpha_S}{\pi} \: xg (x, k^2) \: - C \: \frac{\alpha_S^2}{R^2 k^2} \: \left [ xg (x, k^2) \right  
]^2, 
\ee 
where the last quadratic term, which originates from shadowing, is simply the right-hand-side  
of eq.~(\ref{eq:a1}). 
 
The GLR equation effectively resums the \lq fan\rq\ diagrams generated by the branching of  
QCD Pomerons, which correspond in the GLR approach to gluonic ladders in the DLLA to  
DGLAP evolution.  In this approach the triple-Pomeron vertex, which couple the ladders, is  
computed in the leading $\ln k^2$ approximation.  The GLR equation has stimulated an  
enormous literature 
[3--18] 
%
connected with shadowing effects in deep inelastic and related  
hard scattering processes.  One of the important results to emerge from these studies is the  
computation of the exact triple-Pomeron vertex beyond leading $\ln k^2$, 
but staying within  
the more appropriate leading $\ln 1/x$ approximation. 
 
The aim of the present study is to take advantage of this precise knowledge of the  
triple-Pomeron vertex in order to perform a quantitative estimate of the gluon shadowing  
effects which can be probed in the low $x$ domain which is accessible at the LHC.  To be  
precise, we start from the solution $f_L$ of the unshadowed linear equation which embodies  
both BFKL and DGLAP evolution, as well as subleading $\ln 1/x$ effects
 \cite{KMS}.  Then we compute the quadratic shadowing  
contribution, $-f_{\rm shad}^{(0)}$, from the solution $f_L$ using the more 
complete triple  
Pomeron vertex.  We resum the shadowing contributions using a simple (1,1) Pad\'{e}-type  
representation 
\be 
\label{eq:a3} 
f \; = \; \frac{f_L}{1 + f_{\rm shad}^{(0)}/f_L}, 
\ee 
and the gluon distribution is then calculated from 
\be 
\label{eq:a4} 
xg (x, Q^2) \; = \; xg (x, k_0^2) \: + \: \int_{k_0^2}^{Q^2} \: \frac{dk^2}{k^2} \: f (x, k^2). 
\ee 
 
The structure of the triple-Pomeron vertex can be extracted from an equation, formulated by  
Kovchegov \cite{K}, for the quantity $N (\mbox{\boldmath $r$}, \mbox{\boldmath $b$},  
Y)$.  $N$ is closely related to the dipole cross section $\sigma (r, Y)$ describing the  
interaction of the $q\bar{q}$ dipole of transverse size $r$ with the proton target.
To be precise 
\be 
\label{eq:a5} 
\sigma (r, Y) \; = \; 2 \: \int \: d^2 b \: N (\mbox{\boldmath $r$}, \mbox{\boldmath $b$}, Y), 
\ee 
where $Y = \ln (1/x)$ and $b$ is the impact parameter for the interaction of the $q\bar{q}$  
dipole with the proton.  Recall that the dipole cross section is given in terms of the  
unintegrated gluon distribution by \cite{BNP}
\be 
\label{eq:a6} 
\sigma (r, Y) \; = \; \frac{8 \alpha_S \pi^2}{N_C} \: \int \: \frac{dk}{k^3} \: \left [ 1 - J_0 (k r)  
\right ] \: f (Y, k^2). 
\ee 
 
In the large $N_C$ limit, the function $N$ satisfies the integral equation 
\cite{KSOL}
\bea 
\label{eq:a7} 
& & N (\mbox{\boldmath $r$}_{01}, \mbox{\boldmath $b$}, Y) \; = \; N_0  
(\mbox{\boldmath $r$}_{01}, \mbox{\boldmath $b$}, Y) \: + \: \frac{\alpha_S N_C}{2 \pi}  
\: \int_0^Y \: dy \left \{ - 2 \ln \: \frac{r_{01}^2}{\rho^2} \: N (\mbox{\boldmath $r$}_{01},  
\mbox{\boldmath $b$}, y) \: + \right . \nonumber \\ 
& & \\ 
& & \quad\quad \int_\rho \left . \frac{d^2 r_2}{\pi} \frac{r_{01}^2}{r_{02}^2 r_{12}^2}  
\left [ 2N \left (r_{02}, \mbox{\boldmath $b$} + \textstyle{\frac{1}{2}} \mbox{\boldmath  
$r$}_{12}, y \right ) \: - \: N \left (\mbox{\boldmath $r$}_{02}, \mbox{\boldmath $b$} +  
\textstyle{\frac{1}{2}} \mbox{\boldmath $r$}_{12}, y \right ) \: N \left ( \mbox{\boldmath  
$r$}_{12}, \mbox{\boldmath $b$} - \textstyle{\frac{1}{2}} \mbox{\boldmath $r$}_{20}, y  
\right ) \right ] \right \}, \nonumber 
\eea 
which is the unfolded version of eq.~(15) of Ref.~\cite{K}\footnote{An
equation similar to (\ref{eq:a7}) can be found in Ref.~\cite{BAL}}.  The 
linear part of this equation  
corresponds to the BFKL equation in dipole transverse coordinate space.  The term containing  
the $\log$ denotes the virtual correction responsible for the Reggeization of the gluon, while  
the linear term under the $dr_2$ integral corresponds to real gluon emission.  $\rho$ is the  
ultraviolet cut-off parameter.  The subscripts 01, 02 and 12 enumerate scattering off  
$q\bar{q}, qg$ and $\bar{q}g$ systems respectively.  The equation resums fan diagrams  
through the quadratic shadowing term. 
 
If we rewrite (\ref{eq:a7}) in terms of the transformed function 
\be 
\label{eq:a8} 
\tilde{N} (\ell, b, Y) \; = \; \int_0^\infty \: \frac{dr}{r} \: J_0 (\ell r) \: N (r, b, Y), 
\ee 
then the shadowing term has a much simpler form 
\be 
\label{eq:a9} 
\tilde{N} (\ell, b, Y) \; = \; \tilde{N}_0 (\ell, b, Y) \: + 
\: \frac{\alpha_S N_C}{\pi} \:  
\int_0^Y \: dy \: \left [K \: \otimes \:  \tilde{N} (\ell, b, y) \: - \: \tilde{N}^2 (\ell, b, y) \right ], 
\ee 
where $K$ is the BFKL kernel in momentum space \cite{BFKL}.  Here we have made the short-distance  
approximation in which we neglect the $\mbox{\boldmath $r$}_{ij}$ terms in comparison to  
$\mbox{\boldmath $b$}$, so that $N$ is only a function of the magnitudes $r$ and $b$,
and $\tilde{N}$ of $\ell$ and $b$. 
 
We may resum the linear BFKL effects and rearrange (\ref{eq:a9}) in the form 
\be 
\label{eq:a10} 
\tilde{N} (\ell, b, Y) \; = \; \tilde{N}_L (\ell, b, Y) \: - \: 
\frac{\alpha_S N_C}{\pi} \:  
\int_0^Y \: dy \: G (Y - y) \: \otimes \: \tilde{N}^2 (\ell, b, y), 
\ee 
where $\tilde{N}_L$ is the solution of the linear part of (\ref{eq:a9}) with the shadowing  
term neglected, and $G$ is the Green's function of the BFKL kernel 
\be 
\label{eq:a11} 
G (Y - y) \; = \; \exp \left ( \frac{\alpha_S N_C}{\pi} \: (Y - y) K \right ). 
\ee 
Eq.~(\ref{eq:a10}) may be solved by iteration.  At large $Y (\equiv \ln 1/x)$ the dominant  
region of integration is $y \sim Y$, where $G \simeq 1$, and so the first iteration gives 
\be 
\label{eq:a12} 
\tilde{N} (\ell, b, Y) \; = \; \tilde{N}_L (\ell, b, Y) \: - \: 
\frac{\alpha_S N_C}{\pi} \:  
\int_0^Y \: dy \: \tilde{N}_L^2 (\ell, b, y). 
\ee 
 
We now assume that the $b$ dependence can be factored out of $\tilde{N}_L$ as a profile  
function $S (b)$ 
\be 
\label{eq:a13} 
N_L (\ell, b, Y) \; = \; S (b) \: n_L (\ell, Y), 
\ee 
where we use the normalisation 
\be 
\label{eq:a14} 
\int \: d^2 b \: S (b) \; = \; 1. 
\ee 
Integrating (\ref{eq:a12}) over $d^2 b$ then gives 
\be 
\label{eq:a15} 
\tilde{n} (\ell, Y) \; = \; \tilde{n}_L (\ell, Y) \: - \: 
\frac{\alpha_S N_C}{ \pi} \: \frac{1}{\pi  
R^2} \: \int_0^Y \: dy \: \tilde{n}_L^2 (\ell, y), 
\ee 
where 
\be 
\label{eq:b15} 
\frac{1}{\pi R^2} \; \equiv \; \int \: d^2 b \: S^2 (b). 
\ee 
 
\begin{figure}\centering
\scalebox{0.7}{\resizebox{\textwidth}{!}{\includegraphics[1cm,1cm][17cm,21cm]{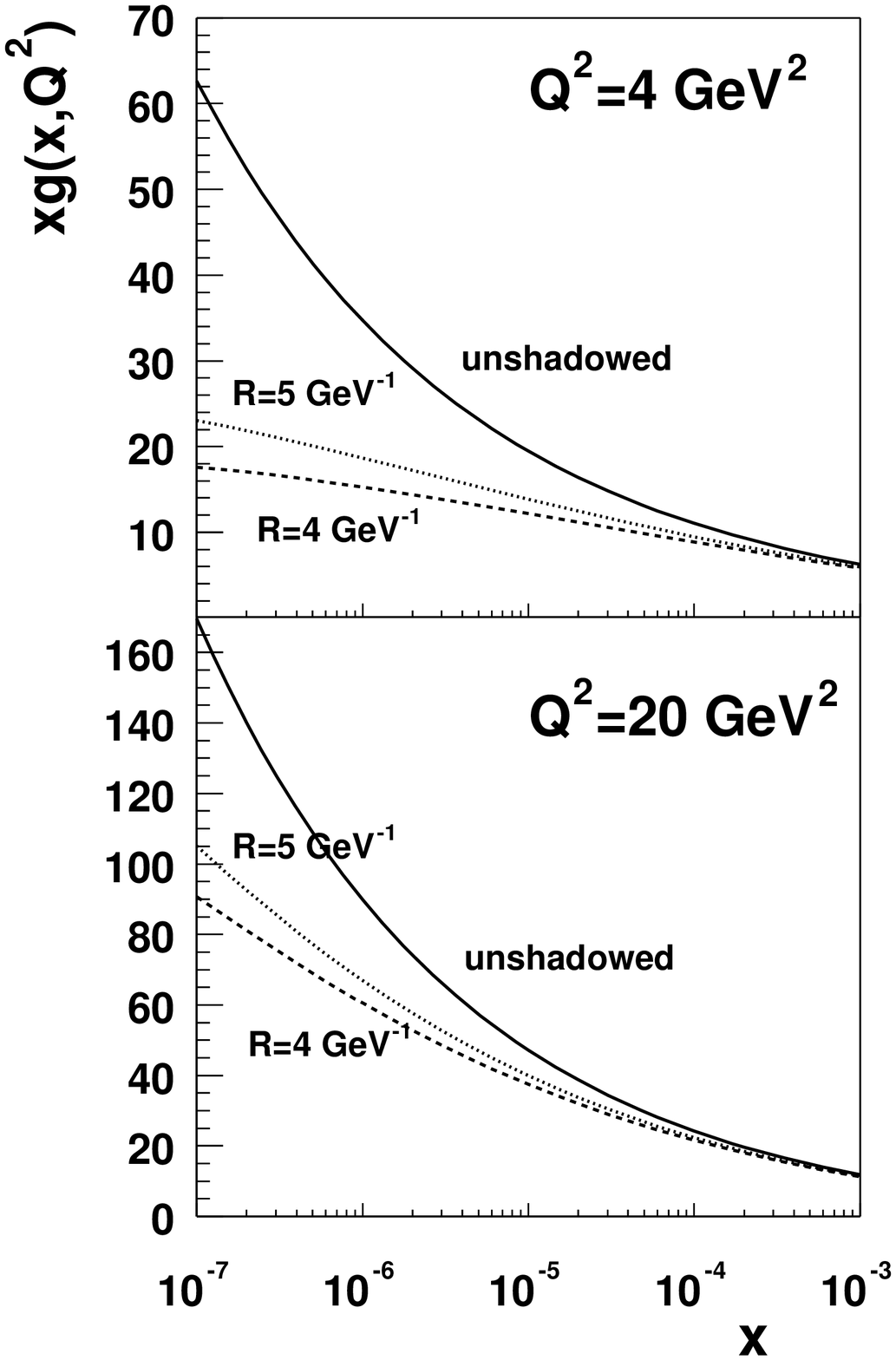}}}
\caption{The effect of shadowing on the integrated gluon distribution $xg(x,Q^2)$, at 
$Q^2 = 4~ {\rm GeV}^2$ and $20~ {\rm GeV}^2$.  The continuous line 
is simply the unshadowed $xg$ obtained from $f_L$ \cite{KMS}.  
The dashed and dotted lines show the result of shadowing with $R=4$ and
$5\ {\rm GeV}^{-1}$
respectively, where the `radius' $R$ is defined in (\ref{eq:b15}) in
terms of the proton profile function, $S(b)$.}
\label{fig:fig1}
\end{figure}

We use (\ref{eq:a6}) and (\ref{eq:a5}) to write (\ref{eq:a15}) in terms of the unintegrated  
gluon distribution.  We obtain  
\be 
\label{eq:a16} 
f (Y, k^2) \; = \; f_L (Y, k^2) \: - \: \frac{\alpha_S^2}{ R^2} \: \left ( 1 \: - \: \frac{d}{d \ln  
k^2} \right )^2 \: k^2 \: \int_0^Y \: dy \: \left [ \int_{k^2}^\infty \: \frac{d \ell^2}{\ell^4} \: \ln  
\left ( \frac{\ell^2}{k^2} \right ) \: f_L (y, \ell^2) \right ]^2, 
\ee 
where we have used the identities 
\bea 
\label{eq:a17} 
& & \int_0^\infty \: \frac{dr}{r} \: J_0 (kr) \: \left [1 - J_0 (\ell r) \right ] \; = \;  
\textstyle{\frac{1}{2}} \: \ln \left ( \displaystyle{\frac{\ell^2}{k^2}} \right ) \: \Theta (\ell^2 -  
k^2), \\ 
& & \nonumber \\ 
\label{eq:a18} 
& & \left ( 1 \: - \: \frac{d}{d \ln k^2} \right )^2 \: k^2 \: 
\int_{k^2}^\infty \: {d\ell^2\over \ell^4} \: \ln \left  
( \frac{\ell^2}{k^2} \right ) \: f (Y, \ell^2) \; = \; f (Y, k^2). 
\eea 
Note that the term in square brackets in (\ref{eq:a16}) is proportional to $n_L (k, y)$.   
Formula (\ref{eq:a16}) is valid in the large $N_C$ limit, but for finite $N_C$ we need to  
multiply the shadowing term by a factor\footnote{Of course the change of the normalisation
of the shadowing term by the factor $N_C^2/(N_C^2 -1)$ does not exhaust possible corrections
beyond the large $N_C$ limit.  One may expect that there will be other contributions
beyond those leading to equation (\ref{eq:a7}).  The general structure of those corrections
is unfortunately not entirely known.  One can however expect that after taking them into
account it may no longer be possible to obtain a closed equation for $N$ since one will
have to introduce other independent and more complicated dynamical quantities as well.  It
should also be observed that the non-linear equation (\ref{eq:a7}) does not contain
possible effects generated by the compound states of more than two reggeised gluons
\cite{ZM}.} $N_C^2/(N_C^2 - 1) = 9/8$.  The second term on  
the right-hand-side of (\ref{eq:a16}) is simply the shadowing contribution $- f_{\rm  
shad}^{(0)}$ of (\ref{eq:a3}).  Recall that (\ref{eq:a3}) represents the (1,1) Pad\'{e}  
approximation of the series whose first two terms are given on the right-hand-side of  
(\ref{eq:a16}).  Moreover, we emphasize again that for the linear term $f_L$ we  use the  
solution of an equation which embodies both BFKL and DGLAP behaviour and which  
contains major sub-leading effects in $\log 1/x$ \cite{KMS}.  In Fig.~\ref{fig:fig1}
 we show the results 
for the integrated gluon $xg (x, Q^2)$ obtained from (\ref{eq:a3}) and (\ref{eq:a4}).  The 
shadowing term $-f_{\rm shad}^{(0)}$ in (\ref{eq:a16}) is computed from the 
unintegrated gluon $f_L$ of Ref.~\cite{KMS},
 assuming a running coupling $\alpha_S (k^2)$. 
 
Several features of the results of Fig.~\ref{fig:fig1} are noteworthy.  First we see,
as expected, the effect of shadowing on $xg\left(x,Q^2\right)$ decreases with increasing
$Q^2$.  Second, with increasing $\ln \left(1/x\right)$, the start of the `turn-over' towards
the saturation limit is evident in the $Q^2=4~{\rm GeV}^2$ curves.  The major uncertainty in
the predictions arises from the choice of the value of $R$, as a consequence of the $1/R^2$
dependence of the shadowing term.  We have chosen values of $R$ that are consistent with
the radius of the proton\footnote{If the gluons were concentrated in `hot-spots' within the
proton, then shadowing effects would, of course, be correspondingly larger.}.  The results of
Fig.~\ref{fig:fig1} show that the effects of shadowing are rather small and difficult to
identify at HERA where, at best, the domain $x\sim 10^{-4} - 10^{-3}$ at $Q^2\sim 5~{\rm GeV}^2$
can be probed.  On the other hand shadowing leads to up to a factor of 2 suppression of $
xg\left(x,Q^2\right)$ in the $Q^2 \sim 5 ~{\rm GeV}^2$ and $x\sim 10^{-6}-10^{-5}$ domain
accessible to the LHC experiments \cite{ADR}.
 
For completeness we summarize how the Kovchegov equation \cite{K}, (\ref{eq:a7}), may be  
reduced to GLR form \cite{GLR}.  We start with (\ref{eq:a6}) and approximate $1 - J_0  
(kr)$ by $(kr)^2/4$, which is valid provided $k^2 \ll 4/r^2$.  Then we obtain 
\be 
\label{eq:a19} 
\sigma (r, Y) \; = \; \frac{\alpha_S \pi^2}{N_C} \: r^2 \: \int^{4/r^2} \: \frac{dk^2}{k^2} \: f  
(Y, k^2), 
\ee 
where the integral can be identified with the integrated gluon $xg (x, 4/r^2)$, where $Y = \ln  
(1/x)$.  Thus, from (\ref{eq:a5}), we have 
\be 
\label{eq:a20} 
\int \: d^2 b \: N (\mbox{\boldmath $r$}, \mbox{\boldmath $b$}, Y) \; \simeq \;  
\frac{\alpha_S \pi^2}{2 N_C} \: r^2 \: xg (x, 4/r^2). 
\ee 
Now if (\ref{eq:a7}) is evaluated in the strongly-ordered approximation $(r_{01}^2 \gg  
r_{02}^2 \sim r_{01}^2)$ it can be shown, using (\ref{eq:a4}), that it reduces to the GLR  
form 
\be 
\label{eq:a21} 
\frac{\partial g (x, Q^2)}{\partial Y \partial \ln (Q^2/\Lambda^2)} \; = \; \frac{N_C  
\alpha_S}{\pi} \: xg (x, Q^2) \: - \: \frac{\alpha_S^2 \pi}{\pi R^2} \: \frac{1}{Q^2} \: \left [xg  
(x, Q^2) \right ]^2. 
\ee 
Comparing with (\ref{eq:a2}) we see that the coefficient\footnote{The corresponding  
coefficient which defined the strength of shadowing in Ref.~\cite{MQ} was 
$C = 4N_C^2/(N_C^2-1)$, which is four times larger in the large $N_C$ limit.} 
$C = 1$. 
 
In summary, we have quantified the size of the shadowing corrections to $xg (x, Q^2)$ using  
a triple-Pomeron vertex which is valid beyond leading $\ln Q^2$, but staying within the  
leading $\ln (1/x)$ approximation.  The corrections are found to be sizeable for $Q^2 \simeq  
4~{\rm GeV}^2$ and $x \simeq 10^{-6}-10^{-5}$, see Fig.~\ref{fig:fig1}.  
This domain may be probed at the LHC  
by observing prompt photon production ($gq \rightarrow \gamma q$) or Drell-Yan  
production both at very large rapidities \cite{ADR}.  Of course the latter process involves a  
convolution to allow for the $g \rightarrow q\bar{q}$ transition, which is required for a  
gluon-initiated reaction; consequently somewhat larger values of the gluon $x$ are probed. 
 
\section*{Acknowledgements} 
 
We thank Albert De Roeck for discussions which prompted this study, 
and Yuri Kovchegov and Genya Levin for illuminating correspondence.  
We are also grateful to Krzysztof Golec-Biernat for critically 
reading the manuscript and important comments, and to Michal Praszalowicz
for illuminating discussions.  
JK thanks the Grey College and the Department of Physics of the University of Durham for their 
warm hospitality.    
This work was supported  
by the UK Particle Physics and Astronomy Research Council (PPARC), and also supported 
by the EU Framework TMR programme, contract FMRX-CT98-0194 (DG 12-MIHT), and the British
Council -- Polish KBN Joint Collaborative Programme.
 

\end{document}